\documentclass[submission,phys]{SciPost} 

\def\apgt{\ {\raise-.5ex\hbox{$\buildrel>\over\sim$}}\ }
\def\aplt{\ {\raise-.5ex\hbox{$\buildrel<\over\sim$}}\ }

\usepackage{graphicx}
\usepackage{subcaption}
\usepackage{amsmath}
\usepackage{tikz}
\usepackage{hyperref}
\usepackage{textcomp}

\newcommand*\aap{A\&A}

\newcommand*\apj{ApJ}
\newcommand*\apjl{ApJ}

\newcommand*\mnras{MNRAS}

\newcommand*\nat{Nature}

\newcommand*\sovast{Soviet~Ast.}

\def\eprint#1#2{#1:#2}
  
\begin{document}

\begin{center}{\Large \textbf{
Lucky planets: how circum-binary planets survive the supernova in one of the inner-binary components
}}\end{center} 

\begin{center}
Fedde Fagginger Auer,\\
Simon Portegies Zwart
\end{center} 

\begin{center}
Leiden Observatory, Leiden University, PO Box 9513, 2300 RA, Leiden, The Netherlands\\
* spz@strw.leidenuniv.nl
\end{center}

\section*{Abstract} {\bf A planet hardly ever survives the supernova
  of the host star in a bound orbit, because mass loss in the
  supernova and the natal kick imparted to the newly formed compact
  object cause the planet to be ejected.  A planet in orbit around a
  binary has a considerably higher probability to survive the
  supernova explosion of one of the inner binary stars.  In those
  cases, the planet most likely remains bound to the companion of the
  exploding star, whereas the compact object is ejected.  We estimate
  this to happen to $\sim 1/33$ the circum-binary planetary systems.
  These planetary orbits tend to be highly eccentric ($e \apgt 0.9$),
  and $\sim 20$\,\% of these planets have retrograde orbits compared
  to their former binary.  The probability that the planet as well as
  the binary (now with a compact object) remains bound is about ten
  times smaller ($\sim 3\cdot 10^{-3}$).  We then expect the Milky way
  Galaxy to host $\aplt 10$ x-ray binaries that are still orbited by a
  planet, and $\aplt 150$ planets that survived in orbit around the
  compact object's companion.  These numbers should be convolved with
  the fraction of massive binaries that is orbited by a planet. }
 
\section{Introduction}
\label{sec:introduction}

Since the discovery of exoplanets around pulsars \cite{Wolszczan1992}
there has been a debate on their origin. Popular scenarios include in
situ formation \cite{Podsiadlowski1993,1993ApJ...412L..33T} or the
dynamical capture of a planet in a dense stellar system
\cite{Sigurdsson2003}.  The possibility of a planet surviving its host
star's supernova is often neglected, because a planet in orbit around
a single exploding star is not expected to survive the supernova
\cite{1993ASPC...36..371P}.  The combination of mass loss in the
supernova explosion \cite{Hills1983} and the natal kick imparted to
the new compact object \cite{Gunn1970,Shklovskii1970} mediates the
survivability of low-mass x-ray binaries
\cite{Kalogera1996,1996MNRAS.281..552T,1998AA...332..173P,2001ASSL..264..355P,2015AA...579A..33V,1998ApJ...506..780B,PortegiesZwart1996},
but planetary orbits are too fragile to survive this process
\cite{Martin2016}.

The survivability of a planet in orbit around a binary, of which one
of the components experiences a supernova is considerably larger than
when the planet directly orbits the exploding star. The lower relative
mass loss in a binary compared to a single star and the dilution of
the velocity kick by dragging along the companion star
\cite{2000A&A...364..563V}, may cause the planet to survive either in
orbit around the binary, or around the original secondary
--non-exploding-- star (in which case the compact object is ejected
from the system).

We calculate the probability that a circum-binary planet survives the
first supernova in a massive binary.  In a first step, we perform
binary population synthesis calculations to determine the orbital
phase-space distribution of the pre-supernova binary system.  We
subsequently add a planet in orbit around the binary and analytically
calculate the supernova's effect on the system to determine the
planet's survivability and post-supernova orbital parameters.

\section{Method}
\label{sec:method}

We approach the problem on the survivability of circum-binary planets
using a combination of techniques. First, we perform a series of
population synthesis calculations for binary stars.  The binaries that
survive until they experience their first supernovae are, in the
second step, provided with a circum-binary planet in a stable orbit,
after which we resolve the supernova explosion. This second step is
calculated analytically for each individual system, and a population
study is carried out through Monte Carlo sampling.

\subsection{Population synthesis calculations of pre-supernova binaries}
\label{subsec:numerical}

Binary evolution is a complicated non-linear problem that is not
easily performed analytically (see however
\cite{1998ApJ...506..780B}).  Therefore we perform this part of the
calculation numerically, using the publicly available and well-tested
binary population-synthesis code {\tt SeBa} \cite{PortegiesZwart1996,
  Toonen2012}.  We adopt the version available in the Astrophysics
Multipurpose Software Environment (AMUSE, \cite{2018araa.book.....P}).
The code takes the metallicities and masses of the two stars ($m_1$
and $m_2$, at zero age) and together with the orbital period and the
eccentricity ($e_i$), the code gives the evolution of these parameters
as a function of time. We adopt Solar metallicity throughout this
study.

There are quite a number of free parameters in a binary population
synthesis code \cite{2014AA...562A..14T}.  The most important ones are
the treatment of non-conservative mass transfer, the amount of angular
momentum per unit mass that is lost by the mass leaving the binary
system ($\beta$), and the common-envelope treatment (often summarized
in the parameters $\alpha \lambda$ and $\gamma$).  We adopt the model
parameters as in \cite{1998AA...332..173P} (their model C, see
table~1, $\alpha\lambda=0.5$, $\beta=6$ and $\gamma=1$), which matches
with the Galactic x-ray binary population.  These parameters are
classically identified with $\alpha$, $\beta$, and $\gamma$, but in
the next section we use the same letters in a different context.

We determine the probability density function for pre-supernova
parameters in phase space through binary population synthesis, and
subsequently, apply a kernel-density estimator to smooth these
distributions and bootstrap the number of systems. In the next step,
we use these smoothed distributions to randomly select pre-supernova
binaries to which we add a planet and subsequently apply the effect of
the supernova.

\subsection{Analytic considerations of the post-supernova system}
\label{subsec:analytical}

The effect of a supernova on a binary system was explored by
\cite{Hills1983}. Later
\cite{Brandt1994,PortegiesZwart1996,Kalogera1996} further studied the
binary's survivability through population synthesis, expanding the
original formulation for \cite{Hills1983} to include elliptical orbits
and a wide range of velocity kicks imparted to the newly formed
compact object.  \cite{Pijloo2012} and \cite{2019MNRAS.484.1506L}
subsequently expanded the formalism to multiple systems by considering
a hierarchy of nested binaries.  We expand on these studies by
adopting a planet around a binary system of which one component
instantaneously loses mass isotropically and receives a velocity kick.

\begin{figure}
\centering
\includegraphics[angle=00, width=0.6\textwidth]{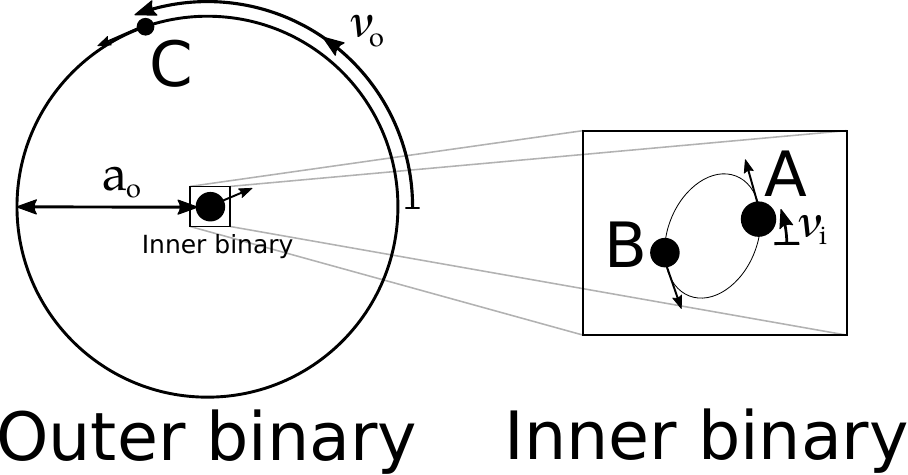}
\caption{The geometry of the hierarchical triple, with inner binary
  composed o stars $A$ and $B$, and the outer planet $C$. Also
  indicated are the outer binary semi-major axis $a_o$ and the true
  anomaly $\nu_i$ and $\nu_o$, for the inner and outer orbit,
  respectively.  }
\label{fig:system_layout}
\end{figure}

In figure \ref{fig:system_layout} we sketch the adopted configuration,
in which an inner binary (star $A$ and star $B$) is orbited by a
planet ($C$)\footnote{Following \cite{PortegiesZwart1996}, we write
  systems by assing the first three letters of the alphabet to the two
  stars and the planet. Parenthesis indicate a bound system, and
  multiple braces a hierachy. The initial triple, for example, then is
  written as $((A, B), C)$, and a possible binary between the seconday
  star and the planet with an unbound primary is written as
  $A, (B, C)$.}.  The properties of the inner (outer) binary are
labeled with subscript $i$ ($o$). The star $A$ undergoes the
supernova.  We further adopt the notation from \cite{Brandt1994}, and
with that the remnant mass $m'_A$ and the average orbital velocity
$v_{\text{orb}}$.
\begin{equation*}
\tilde{m} = \frac{m_A+m_B}{m'_A+m_B}, \hspace{0.5cm} \tilde{v} =
\frac{|\mathbf{v}_{\text{kick}}|}{v_{\text{orb}}}, \hspace{0.5cm}
\mu_B = \frac{m_B}{m_A+m_B}.
\end{equation*}
For a circular binary we have the expression for the orbital energy
after the supernova:
\begin{equation*}
\begin{split}
E' &= -\frac{Gm'_Am_B}{2a}\Delta, \\ \Delta &=
\frac{2a}{|\mathbf{r}-\delta\mathbf{r}|} - \tilde{m}(1+\tilde{v}^2 -
2\tilde{v}\cos{(\theta)}\cos{(\phi)}).
\end{split}
\end{equation*}
Here $\theta,\phi$ give the kick direction (we define $\theta=\phi=0$
to be a kick anti-parallel to orbital motion, and therefore have a
minus sign difference with \cite{Brandt1994}), $a$ is the semi-major
axis of the binary before the supernova, and $\mathbf{r}$ the relative
position vector between the binary components. The shift
$\delta\mathbf{r}$ allows for an instantaneous displacement of the
center of mass of the inner binary. This displacement of the inner
binary results from supernova shell when it passes the companion star
(see \cite{Pijloo2012}).  A bound orbit requires $E'<0$ and therefore
$\Delta > 0$. This inequality gives the physical requirements on the
mass loss and velocity kick.  Combining this with the method of
\cite{Pijloo2012} we can derive a similar formula for the outer binary
in the hierarchical triple.  We assume that both binaries are
initially circular and co-planar (see figure
\ref{fig:system_layout}). We select the inner binary's true anomaly
$\nu_i=0$ by an appropriate rotation of our coordinate system and
assume $m_C \ll m_A,m_B$. We define a $\Delta_{i},\Delta_{o}$ for both
binaries. The supernova in the inner binary leads to an `effective'
supernova in the outer binary through an instantaneous change in the
mass, position, and velocity of the inner binary center of mass:
\begin{equation*}
\begin{split}
\tilde{m}_o &= \frac{(m_A+m_B)+m_C}{(m'_A+m_B)+m_C} \approx
\frac{m_A+m_B}{m'_A+m_B} = \tilde{m}_i \equiv
\tilde{m},\\ \delta\mathbf{r}_{\text{COM}} &= -\mu_B(\tilde{m}-1)a_i
\hat{\mathbf{r}}_{i},\\ \delta\dot{\mathbf{r}}_{\text{COM}} &=
v_{\text{orb},i}[-\mu_B(\tilde{m}-1)\hat{\dot{\mathbf{r}}}_{i} +
  (1-\tilde{m}\mu_B)\tilde{v}_i\hat{\mathbf{v}}_{\text{kick}}].
\end{split}
\end{equation*}
We define auxiliary variables:
\begin{equation*}
\begin{split}
l &= a_o/a_i,\\
K_1 &= \mu_B(\tilde{m}-1) = \mu_B(m_A-m'_A)/(m'_A+m_B),\\
K_2 &= 1-\tilde{m}\mu_B = m'_A/(m'_A+m_B),\\
\alpha &= \left(1+(K_1 l^{-1})^2 + 2 K_1 l^{-1}\cos{(\nu_o)}\right)^{-1/2},\\
\beta &= \left(1 + K_1^2l + 2K_1\sqrt{l}\cos{(\nu_o)}\right)^{1/2},\\
\gamma &= K_2\sqrt{l},\\
\phi_0 &= \arctan{\left(\frac{\sin{(\nu_o)}}{\cos{(\nu_o)}+K_1\sqrt{l}}\right)},\\
&\hspace{1cm} + \frac{\pi}{2}\left(1-\text{sgn}\left(\cos{(\nu_o)}+K_1\sqrt{l}\right)\right)
\end{split}
\end{equation*}
Letting $\tilde{v} = \tilde{v}_i$, this leads to expressions for
$\Delta_{i,o}$:
\begin{equation*}
\begin{split}
\Delta_i &= 2 - \tilde{m}[1+\tilde{v}^2 - 2\tilde{v}\cos{(\theta)}\cos{(\phi)}],\\
\Delta_o &= 2\alpha - \tilde{m}[\beta^2 + (\gamma\tilde{v})^2 + 2\beta\gamma\tilde{v}\cos{(\theta)}\cos{(\phi-\phi_0)}].
\end{split}
\end{equation*}

Note that each of the terms in $\Delta_o$ are re-scaled versions of
similar terms in $\Delta_i$.  Each of the factors
$\alpha,\beta,\gamma$ gives the effect of the supernova in the inner
binary on the outer binary. The effect of kick direction compared to
the planet's true anomaly on the outer binary survivability is given
by $\phi_0$.  We can obtain limits on the magnitude of the kick and
its direction by following \cite{Brandt1994}. A new constraint for the
triple system is the existence of a maximum $\tilde{m}$.  From the
inequalities $\Delta_{i} > 0,\Delta_{o} > 0$ we get requirements on
the kick magnitude. Specifically, $\tilde{v}$ must be in the intervals
for the inner and outer orbits:
\begin{equation*}
\left[1-\sqrt{\frac{2}{\tilde{m}}},
  1+\sqrt{\frac{2}{\tilde{m}}}\right] \hspace{0.05cm}
\text{and} \hspace{0.05cm}
\left[\frac{\beta}{\gamma}-\sqrt{\frac{2\alpha}{\tilde{m}\gamma^2}},
  \frac{\beta}{\gamma}+\sqrt{\frac{2\alpha}{\tilde{m}\gamma^2}}\right].
\end{equation*}

\begin{figure}
\centering
\includegraphics[width=0.6\textwidth]{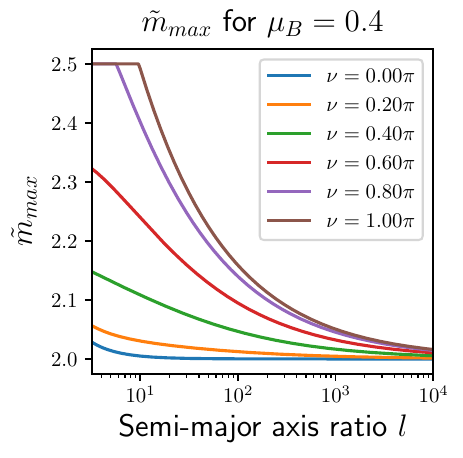}
\caption{Maximum relative mass loss
  $\tilde{m}_\text{max} = (m_A+m_B)/(m'_{A,\text{min}}+m_B)$ as a
  function of semi-major axis ratio $l=a_o/a_i$ for various outer
  binary true anomalies.}
\label{fig:appendix:massloss}
\end{figure}

In the limit $m'_A\to 0$ the value of $\gamma^{-1}$ diverges, and the
lower bound of the second interval may exceed the upper bound of the
first interval (depending on its sign). In such a case, the system
cannot remain bound, leading to a maximum allowed mass loss
$\tilde{m}_{\text{max}}$. This value is limited by the constraint
$m'_A\geq 0$, which sets $\tilde{m}\leq\tilde{m}_{\text{max}}\leq
\mu_B^{-1}$. This is plotted in figure \ref{fig:appendix:massloss}.

\begin{figure}
\centering
\includegraphics[width=0.6\textwidth]{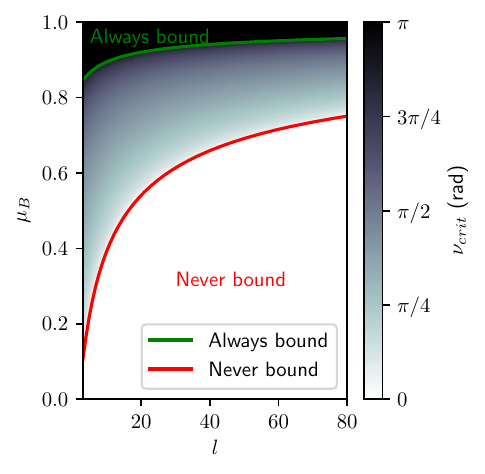}
\caption{Boundaries in $\mu_B,l,\nu_o$ phase space for which the
  $(B, C)$ subsystem remains bound when the remnant star $A$ is
  ejected from the system. The system $(B, C)$ remains bound if
  $\nu_o\in(\pi-\nu_\text{crit},\pi+\nu_\text{crit})$. }
\label{fig:appendix:BC_bound}
\end{figure}

The survivability of the $(B, C)$ as a bound subsystem with $A$
ejected can be investigated by letting $m'_A\to 0$ in $\Delta_o$. 
Examining the limits $\Delta_{(B, C)}=0$ gives the
boundaries in $\mu_B,l,\nu_o$ phase space where $(B, C)$ remains
bound. This is plotted in figure \ref{fig:appendix:BC_bound}. For each
pair $\mu_B,l$ we maximize $\Delta_{(B, C)}$ by picking
$\nu_o=\pi$. Using that $\cos{(x)}$ is an even function and that we maximize
$\Delta_{(B,C)}$ by setting $\nu_o=\pi$, we can define a
$\nu_\text{crit}(\mu_B,l)$ so that $\Delta_{(B, C)}>0$ if
$\nu_o\in(\pi-\nu_\text{crit},\pi+\nu_\text{crit})$. For a uniformly
random value of $\nu_o$ at the time of the supernova the probability of the
planet remaining bound is $\nu_\text{crit}/\pi$.

\section{Results}
\label{sec:results}

Having set out the framework for determining the pre-supernova binary
properties, and describing the effect of the supernova on the triple,
we now combine both to study the survivability of a circum-binary
planet. 

\subsection{Specific choice of initial conditions}

We initialize zero-age main-sequence binaries and evolve them using
population synthesis (see section\,\ref{subsec:numerical}).  In
table\,\ref{tab:priors1} we give the initial conditions.

\begin{table}
\centering
\caption{Initial parameter distributions and value ranges for the
  inner eccentricity $e_i$, mass $m_1$, mass ratio $m_2/m_1$, and
  orbital period $P$.  the primary mass-function and eccentricity
  limits are adopted from \cite{Kroupa2001} and \cite{Kiminki2012},
  respectively.  }
\begin{tabular} {lll}
\hline
\noalign{\smallskip}
Parameter & Distribution shape & Value range \\
\hline
$m_1[\text{M}_\odot]$   & $(m_1[\text{M}_\odot])^{-2.3}$ & $[10, 100]$ \\
$m_2/m_1$              & $(m_2/m_1)^{0.1}$ & $[6\cdot10^{-3}, 1]$ \\
$\log_{10}(P[\text{d}])$& $\log_{10}(P[\text{d}])^{0.2}$ & $[0,4]$ \\
$e_i$ & $e_i^{-0.6}$    & $[10^{-4}, 0.9]$ \\
\hline
\end{tabular}
\label{tab:priors1} 
\end{table}

Binaries that merge or become unbound before the supernova occurs are
discarded.  We subsequently bootstrap the number of systems by
training an \texttt{sklearn} kernel density estimator on $\ln{(m_A)}$,
$\ln{(m_B)}$,
$\ln{(a_i)}, \ln{(e_i/(1-e_i)}, \ln{(a_{i,\text{max}})}$, and the
logarithm of the remnant mass. This ensures that the values drawn from
this distribution are positive and $e_i$ is in the range $(0,1)$. Here
$a_{i,\text{max}}$ is the largest semi-major axis encountered during
the evolution of the inner binary and $e_i$ the inner binary's
eccentricity.

The mass of a black hole is determined by {\tt SeBa}, but for the
neutron-star mass we adopt a Gaussian distribution with mean $1.325$
$M_\odot$ and standard deviation of $0.1125$ $M_\odot$ independent of
the progenitor mass (figure 2 in \cite{Kiziltan2013}).  The other
parameters for the supernova kick are drawn randomly from the
distribution functions presented in table \ref{tab:priors2}.

The inner binary is generally circularized due to mass transfer before
the supernova occurs. In some wide binaries (those with orbital
periods $\apgt 20$\,yr), however, this may not be the case. These wide
and eccentric binaries tend to be very fragile for the supernova, in
particular the planet in an even wider and stable orbit around such
binaries are susceptible to being ionized as a result of the supernova
explosion. In some cases, however, the combination of mass loss and
the natal supernova kick may cause the planet to remain bound. To
accommodate this possibility, we relax the assumption of a circular
inner binary just before the supernova.  In our Monte Carlo sampling,
we randomly select the true anomaly of the inner binary to determine
the effect of the supernova.  A uniformly distributed true anomaly
introduces a bias towards smaller separations and consequently higher
orbital velocities in comparison with adopting a uniform distribution
in the mean anomaly.  We correct for this effect by determining, for
each binary, the probability distribution of mean anomalies by
weighting the simulation results, and scale the supernova survival
probability accordingly.

\begin{table}
\centering
\caption{Distribution parameters for the hierarchical triple and
  velocity kick. Here $e$ refers to eccentricity, $i$ to inclination,
  and $\omega$ to argument of periapsis. The parameter
  $a_{i,\text{max}}$ is the maximum inner binary semi-major axis found
  during the \texttt{SeBa} evolution.  The lower and upper limits to
  the semi-major axis are from \cite{Hobbs2005} and
  \cite{Zhuchkov2010}, respectively.  }
\begin{tabular} {llp{2.3cm}}
\hline
\noalign{\smallskip}
Parameter & Distribution shape & Value range /parameters \\
\hline
$a_o/a_{i,\text{max}}$ & $(a_o/a_{i,\text{max}})^{-1}$ & $[3.23, 1000]$ \\ 
$|\mathbf{v}_{\text{kick}}|[\text{km/s}]$ & Maxwellian & $\sigma=265$ \\
Kick direction & Uniform & Sphere \\
$m_C[M_\odot]$ & Uniform & $[10^{-5}, 5\cdot 10^{-3}]$ \\
$e_o$ & - & 0 \\
$i_o$ & - & 0 \\
$i_i$ & - & 0 \\
$\nu_o$ & Uniform & $[0, 2\pi]$ \\
$\nu_i$ & Uniform & $[0,2\pi]$ \\
$\omega_i$ & Uniform & $[0,2\pi]$ \\
\hline
\end{tabular}
\label{tab:priors2}
\end{table} 

\subsection{Evolution of the inner binary}

After having determined the initial parameter space, we evolve a total
of $4.8\times 10^4$ zero age binaries up to the moment of the first
supernova.

A fraction of $0.14$ experienced a supernova resulting in a black
hole, $0.08$ produced a neutron star in a supernova explosion, $0.03$
left no remnant after the supernova; the rest either resulted in a
merger, an unbound system or did not experience a supernova.  We do not
discuss the evolution of these systems here, but adopt the eventual
distributions of the orbital parameters of the surviving binaries to
further explore the possibility that a circum-binary planet survives
the supernova.

\begin{figure}
\centering
\subcaptionbox{Pre-supernova masses for binaries with a neutron star.}%
[.45\textwidth]{\includegraphics[width=0.5\textwidth]{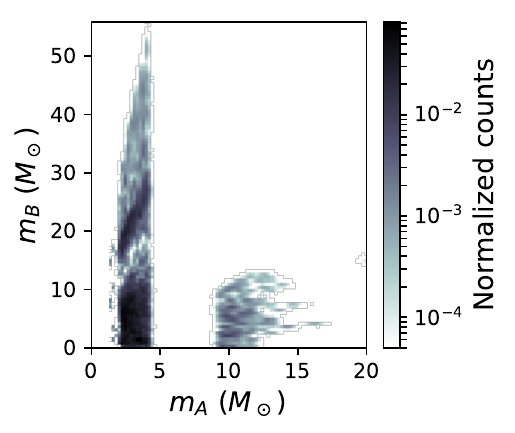}}
\subcaptionbox{Pre-supernova masses for binaries with a black hole.}%
[.45\textwidth]{\includegraphics[width=0.5\textwidth]{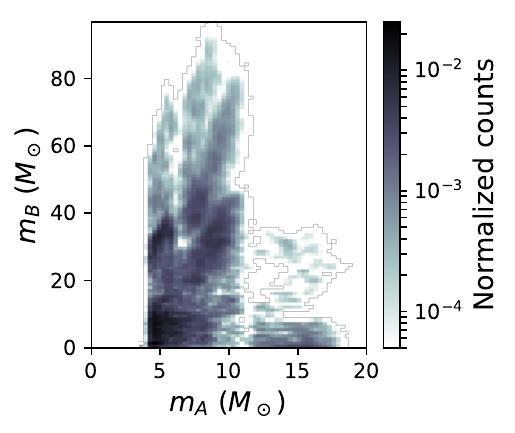}}
\subcaptionbox{Semi-major axis versus eccentricity for neutron-star binaries directly after the supernova}%
[.45\textwidth]{\includegraphics[width=0.5\textwidth]{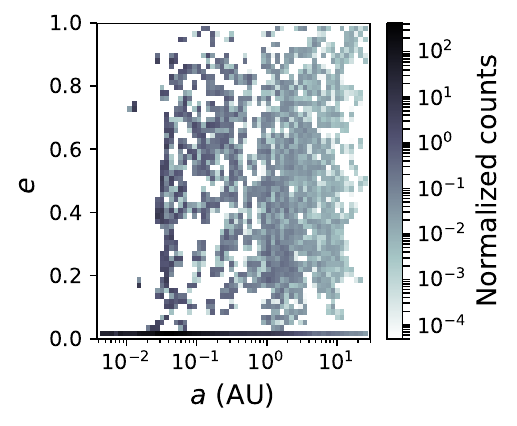}}
\subcaptionbox{Semi-major axis versus eccentricity for black-hole binaries directly after the supernov.}%
[.45\textwidth]{\includegraphics[width=0.5\textwidth]{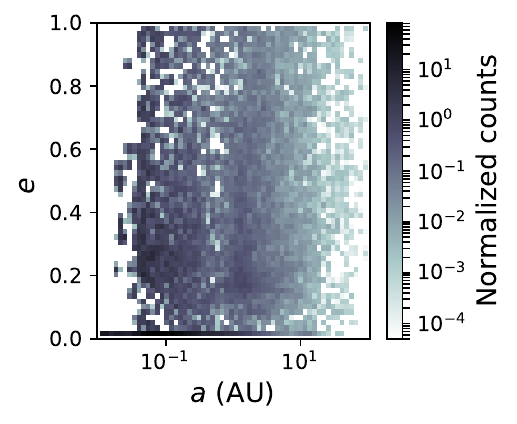}}
\caption{Distribution of the parameters for the inner binaries just
  before and after the supernova for both neutron stars (left) and
  black holes (right).}
\label{fig:SeBa_results}
\end{figure}

Figure \ref{fig:SeBa_results} shows the parameter distributions for
the simulations leaving a remnant.  Most of the supernova progenitors
have a mass $m_A < 10\,M_{\odot}$, while there are secondary masses as
high as $80\,M_{\odot}$. Most of the supernov\ae\, are stripped
core-collapse of type Ibc that naturally result from the mass transfer
during the system evolution \cite{Tauris2013}.  Most inner binaries
have semi-major axes $<1$ au.

\subsection{Survivability of the planet after the supernova}

In order to study the probability that a planet survives in orbit
throughout the supernova, we add a planet, with a mass chosen
uniformly between $10^{-5}$\,M$_\odot$\, and
$5\times10^{-3}$\,M$_\odot$ and orbital separation between
$a_o = 3.23\,a_{i, \text{max}}$ to $1000\,a_{i, \text{max}}$ (flat in
$\log(a_o)$).  Here $a_{i, \text{max}}$ is the largest semi-major axis
the inner binary reaches during its pre-supernova evolution.  The
inner limit is chosen to assure that the planet would remain stable
throughout the evolution of the inner binary
(\cite{1999ASIC..522..385M,2018MNRAS.476..830M}, and see
\cite{Zhuchkov2010} for a more empirical characterization).  Mass lost
from non-conservative mass transfer in the inner binary or by the wind
of one of the components will have driven the planet further out
\cite{2011MNRAS.417.2104V}, but we ignore that here. The planets are
chosen to have prograde circular orbits in the plane of the
pre-supernova binary system.

The supernova is simulated applying an instantaneous mass loss and
change in the velocity vector of star $A$. This leads to a sudden
change in the center of mass and velocity of the inner binary, which
has the effective of a (diluted) supernova in the outer orbit, as
discussed in \cite{Pijloo2012}.

Neutron stars and black holes acquire a velocity kick upon their
formation in a supernova.  This kick's magnitude and direction are
crucial for the survivability of the binary system and important for
the orbital parameters when the binary remanis bound.  We adopt the
distribution of pulsar kicks from \cite{Hobbs2005}, which is described
by a Maxwellian distribution with a dispersion of $\sigma=265$
km/s. This distribution appears consistent with population statistics
of x-ray binaries in the Milky Way \cite{1998AA...332..173P}.  Black
holes are expected to receive a lower kick velocity, which we address
by applying a momentum-kick.

The supernova in a triple then results in one of the following:
\begin{itemize}
\item The entire triple becomes unbound: $A, B, C$,
\item The triple becomes dynamically unstable,
\item The inner binary remains bound, but the planet escapes: $(A, B), C$,
\item $A$ and $C$ remain bound, but $B$ escapes: $(A, C), B$,
\item $B$ and $C$ remain bound, but $A$ escapes: $A, (B, C)$,
\item The entire triple remains bound: $((A, B), C)$,
\end{itemize}
We look at every possible combination of $A, B$ and $C$ and determine
the distribution of the corresponding orbital elements. We recognize
two distinct dynamically unstable configurations: 1) the entire triple
remains bound but violates the stability criterion
\cite{Zhuchkov2010}, and 2) the is no clear hierarcy in the surviving
triple. In the latter case we often find that The planet is bound to
both stars, but the two stars are not bound in a binary.

Using the earlier mentioned binary population synthesis results, we
generate $4\times 10^{6}$ pre-supernova systems with a neutron star
and $4\times 10^{6}$ with a black hole, each of which with a
circum-binary planet.

\begin{figure*}
\subcaptionbox{Black hole remnants.}%
[1.0\textwidth]{\includegraphics[width=0.8\textwidth]{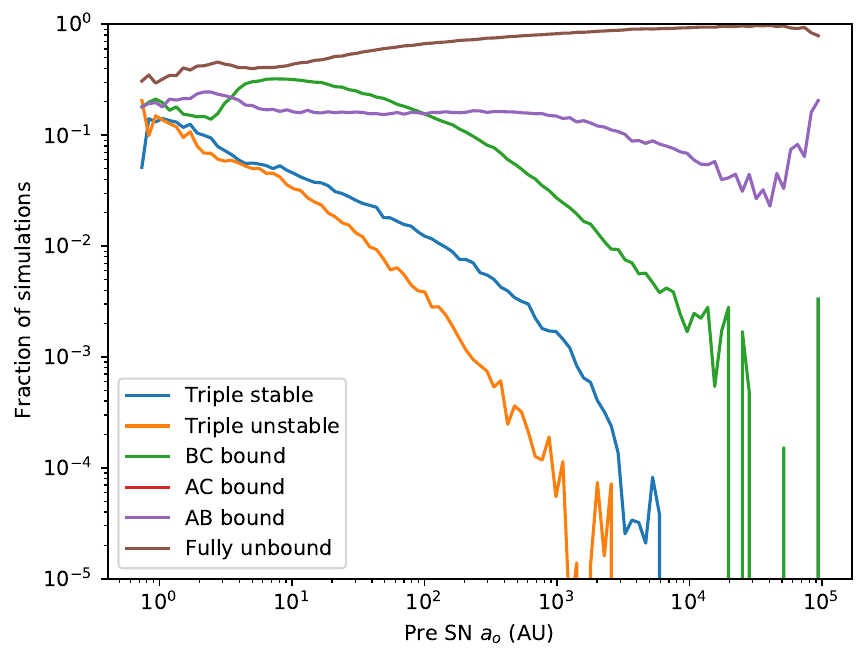}}
\subcaptionbox{Neutron star remnants.}%
[1.0\textwidth]{\includegraphics[width=0.8\textwidth]{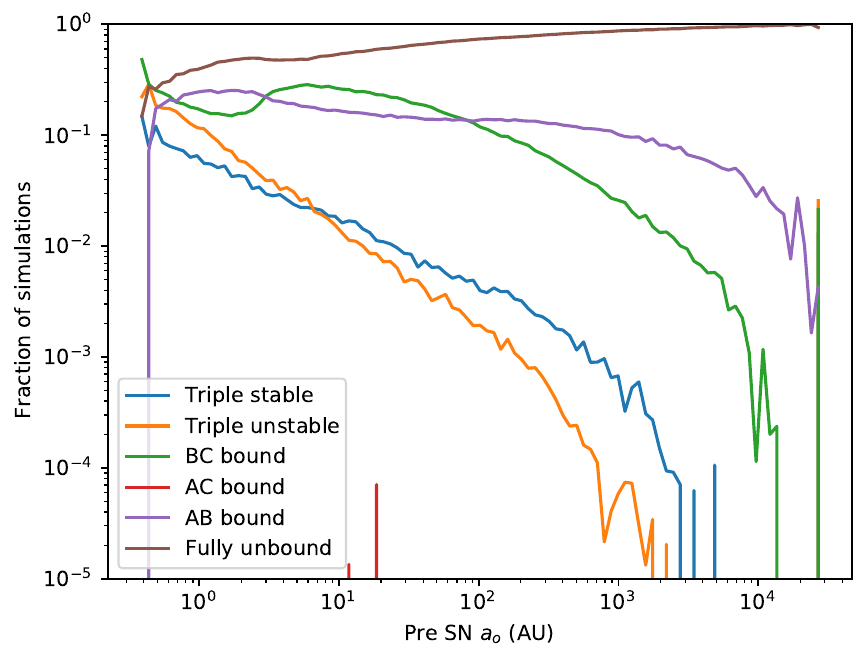}}
\caption{Fraction of simulations per possible outcome of the supernova
  as a function of the outer binary semi-major axis $a_o$ before the
  supernova. The fraction of simulations in these figures is
  calculated as the ratio of two histograms of $a_o$ with logarithmic
  bins, comparing the number of simulation with a specific outcome to
  the total number of simulations in each bin. The `triple unstable'
  category refers to systems which remain fully bound but for which
  the outer planet orbit is unstable \cite{1999ASIC..522..385M}. }
\label{fig:pbound}
\end{figure*}

In figure \ref{fig:pbound}, we show, as a function of $a_o$, the
fraction of simulations that experience a supernova (see
\ref{subsec:numerical}). The increasing noise for $a_o>10^3$ au is due
to the smaller number of simulations in that region of parameter space
(note that $a_0$ was initially distributed randomly on a $\log$
scale). Another interesting feature is that for $a_o \aplt 10^2$ au
the probability of $(B, C)$ to remain bound is larger than for
$(A, B)$ to stay bound. The majority of triples that survive the
supernova are dynamically stable, which is a direct consequence of the
adapted stable conditions before the supernova.

In only 3 of the $8\times 10^6$ simulations, the planet remains bound
to the exploding star, leading to a binary composed of $(A, C)$. For
this to happen, the remnant must receive a kick in a narrow cone and
with the velocity similar to the planets orbital speed but away from
the companion star so that it can pick-up the planet on its escaping
trajectory. Alternatively, the compact object escapes, leaving it
companion star $B$ bound to the planet. Both processes are rather
improbable, as expected, which is consistent with the small
probability in our simulations.

\begin{figure}
  \center
  \includegraphics[angle=00, width=0.6\textwidth]{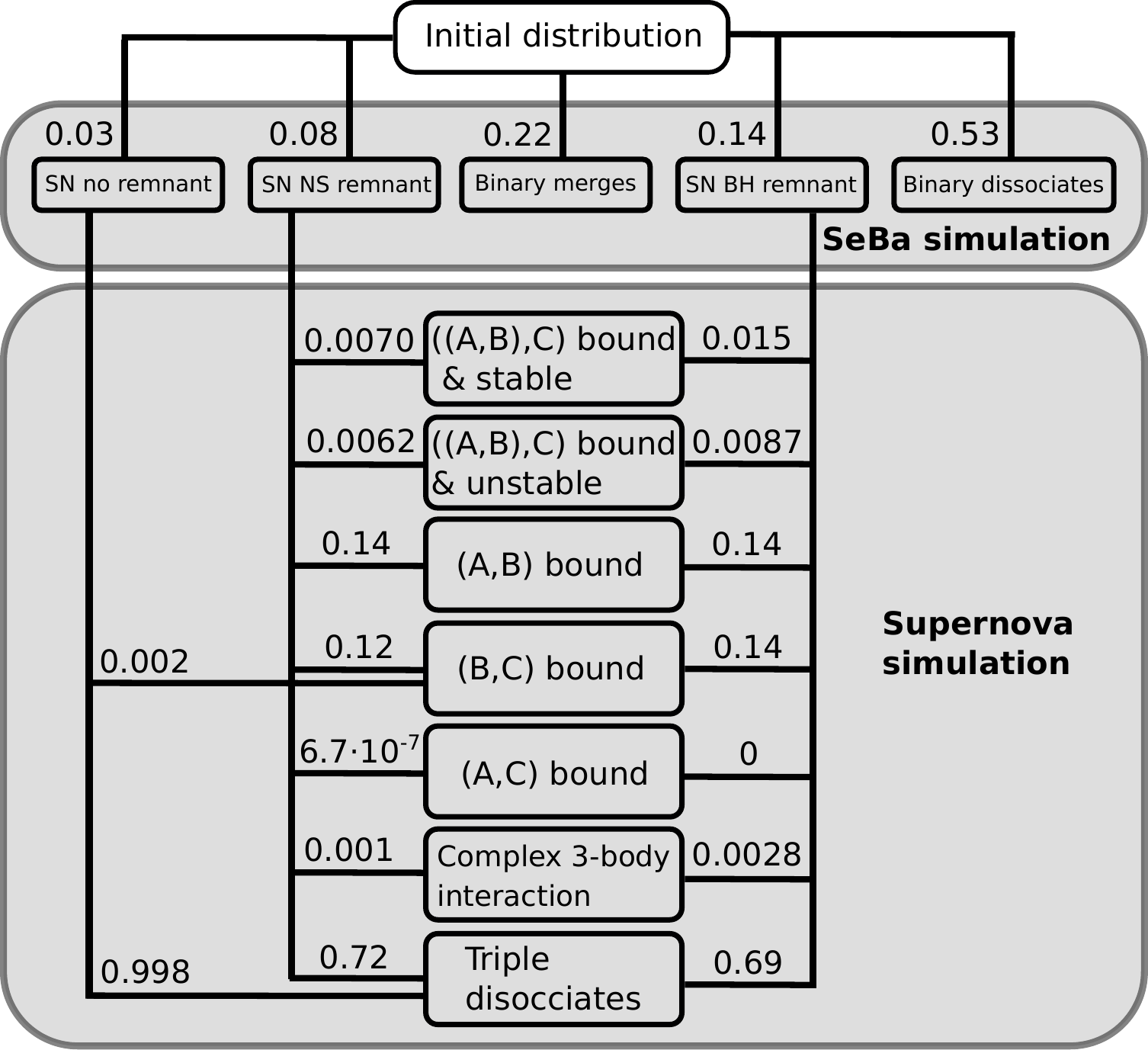}
  \caption{Branching ratios for each possible outcome for the triple
    system, averaged over all input parameters. The upper branches
    give the results of the binary population synthesis, and the lower
    branch gives the effect of the supernova on the resulting
    systems.}
\label{fig:ptree}
\end{figure}

In figure \ref{fig:ptree} we present the branching ratios of all
occurrences in our simulations.  Following the appropriate branches,
the total probability for the triple to survive is $2.7\cdot 10^{-3}$,
and the probability that $(B, C)$ remains bound is $3\cdot 10^{-2}$.
The currently detected number of x-ray binaries is around 300
\cite{Liu2006, Liu2007}, with an estimates for the Galactic population
of $\sim1300$ \cite{Corral-Santana2016} to $10^4$
\cite{Yungelson2006}. With these last two estimates, the expected
number of x-ray binaries that after the supernova that still host a
circum-binary planet is $\aplt 10$: the Galaxy may host a few x-ray
binaries with a circum-binary planet that survived its host's
supernova. There is a comparable probability for the entire triple to
remain bound but with an unstable orbit for the planet. In the latter
case, the planet may either collide with one of the stars or be
ejected to become a rogue planet.

\begin{figure}
\centering
\subcaptionbox{Black hole remnants.}%
[1.0\textwidth]{\includegraphics[width=0.5\textwidth]{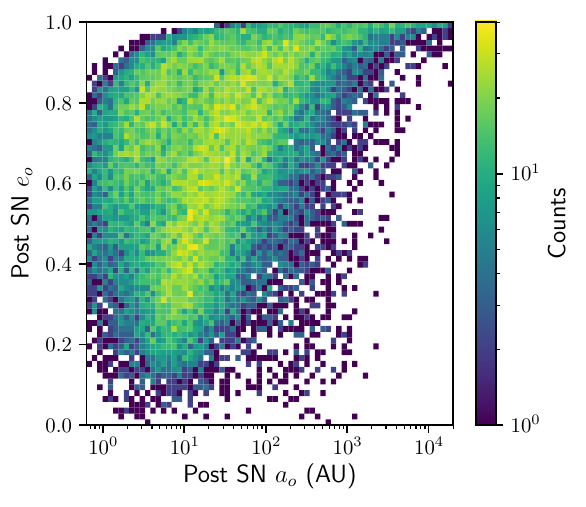}}
\subcaptionbox{Neutron star remnants.}%
[1.0\textwidth]{\includegraphics[width=0.5\textwidth]{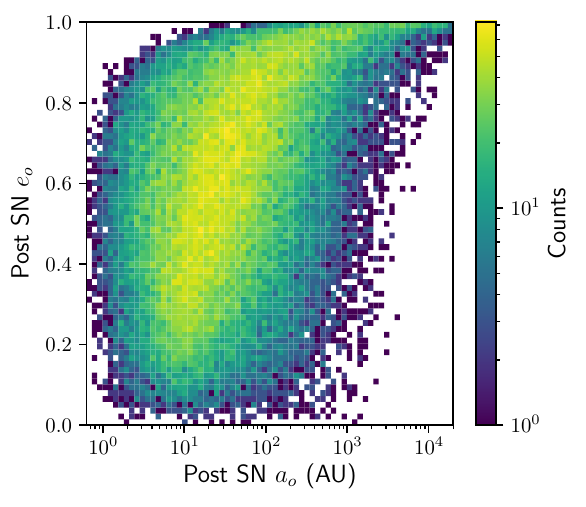}}
\subcaptionbox{Case for which the exploding star is ejected.}%
[1.0\textwidth]{\includegraphics[width=0.5\textwidth]{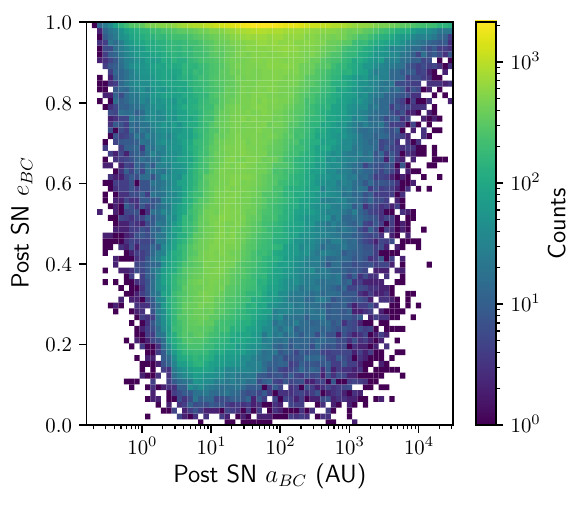}}
\caption{Distribution for semi-major axis and eccentricity for the
  orbiting planet that survive the supernova of one of the binary
  components. The panels a) and b) give the orbital elements for the
  circum-binary planet with a black hole and a neutron star,
  respectively.  Panel c) gives the orbital parameters for the planet
  in orbit around star $B$.}
\label{fig:orbits}
\end{figure}

In figure \ref{fig:orbits} we present the distribution of the
semi-major axis and eccentricity of the surviving planet's orbit. The
planet then either continues to orbit the inner binary, or it orbits
the original secondary star.  Mass loss in the inner binary generally
induces a considerable eccentricity in the final orbit.  Surviving
planets therefore typically have highly eccentric orbits, with a wide
range of semi-major axes.

There is an absence of stable triples with semi-major axes larger than
$10^3$ au and low eccentricities, because the number of planets with a
pre-supernova semi-major axis $>10^3$\,au is low, and the
post-supernova periapsis of the planet can not exceed the
pre-supernova apoapsis.  No planets are found with tight $\ll 1$\,au
orbits because these systems tend to be dynamically unstable, and
wider, up to $\sim 10$\,au orbits tend to have high eccentricities.
Most of the orbits in the $(B, C)$ systems that remain bound have
$e_{(B, C)}\uparrow 1$ due to the extreme mass loss in the triple.  We
predict systems that formed this way to have highly eccentric
planetary orbits, with no strong constraints on the semi-major axis.

\begin{figure}
\centering
\includegraphics[angle=00, width=0.7\textwidth]{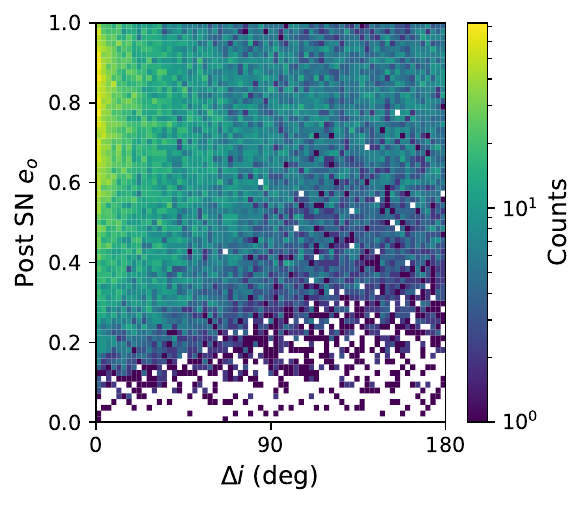}
\caption{Relative inclination of the planet orbit with respect to the
  post-supernova binary, presented in the case where the supernova
  resulted in a neutron star. $\Delta i = 0$ indicates that the planet
  still orbits in the plane of the post-supernova binary system. The
  fraction of retrograde orbits is between 0.148 (for binaries with a
  black hole) and 0.277 (for binaries with a neutron star).}
\label{fig:inclination}
\end{figure}

In figure\,\ref{fig:inclination} we present the distribution of
relative inclinations for the planetary orbits with respect to the
post-supernova binaries.  In this case, we used the results for the
neutron stars, but the black hole distribution is similar. The
fraction of retrograde orbits for circum-binary planets with a neutron
star is 0.277 compared to 0.148 for binaries with a black hole. A
fraction of 0.206 of the planets around a companion of the exploding
star in $(B, C)$ has a retrograde orbit.

\section{Observational implications}
When both stars and the planet remain bound, the inner binary hosting
the compact object continues to evolve into an x-ray binary.  In that
case, the circum-binary planet may be observed in the x-ray binary
phase or around a binary millisecond pulsar once the neutron star has
been spun-up. In principle, it is even possible that the planet
survives a second supernova explosion, in which case the planet
remains bound to the inner compact-object binary.  This requires the
triple to survive and remain stable through both supernov\ae. The
probability of this to happen seems small
($\aplt (0.14 \times 0.015)^2 \sim 4.4 \cdot 10^{-6}$, for a black
hole binary and more than an order of magnitude smaller for neutron
star binaries,
($\aplt (0.08 \times 0.00070)^2 \sim 3.1 \cdot 10^{-7}$, see
figure\,\ref{fig:ptree}).  When the compact object is ejected, the planet
may still survive in a relatively wide and eccentric orbit around the
original secondary star. Such systems may be recognizable for their
curious planetary orbit and the mass-transfer affected stellar host.

The orbits of the surviving planets are wide, with semi-major axes
ranging from $\sim 1$\,au to well over $10^3$\,au, and over the entire
range of eccentricities, but skewed to $e \apgt 0.2$ orbits.  The
majority of systems in which the planet remains bound to the companion
of the exploding star tends to have very highly eccentricities
$e \apgt 0.9$ orbits (but skewed to $e \sim 1)$. Such a system may
lead to a merger between the planet and its newly acquired companion
star within a few years after the supernova.  For the surviving
systems, the discovery of a massive star with a planet in a wide ($10$
to $10^3$\,au) and eccentric ($\apgt 0.9$) orbit may be a signature of
a supernova survival.  These planets will have experienced a nearby
supernova, possibly obliterating their atmosphere, or at the least
enriching it with a healthy radioactive mix of heavy decay-product of
the supernova blastwave.

A considerable fraction of the pre-supernova systems have
eccentricities $e > 0.1$ ($0.24$ of systems with a neutron star and
$0.32$ of systems with a black hole). A uniformly random true anomaly,
as we adopted in our analysis, is biased to lower separations and
higher orbital velocities compared to a mean anomaly selected randomly
from a uniform distribution for eccentric systems. This impacts the
survivability of the system, and introduces a bias in our results. We
correct for this by calculating how much the mean anomaly of a
specific system is overcounted (compared to a uniformly random mean
anomaly), and scaling the effect of that system on our results by the
inverse of this.

\section{Conclusion}

\label{sec:conclusion}

We simulated a population of massive zero-age binaries up to the
moment of the first supernova.  The surviving binaries were equipped
with a circum-binary planet to determine the probability distributions
of the planets' orbital parameters of the surviving binaries. We did
this by analytically investigating limiting factors on the planet's
survival and through Monte Carlo sampling. The analytic expressions
for the amount of mass lost to assure that the planet remains bound
and the probability of remaining bound are presented in figures
\ref{fig:appendix:massloss} and \ref{fig:appendix:BC_bound},
respectively. The resulting numerically calculated survivability for
the planet is presented in figure\,\ref{fig:pbound}.

From the total population of massive Galactic binaries that evolve
into x-ray binaries, we predict a fraction of $3\cdot 10^{-3}$ to keep
its circum-binary planet. This fraction should be perceived as an
upper limit, because we assumed 100\% triplicity.  Interestingly
enough, the probability for the planet to remain bound to the
exploding star's companion is $11$ ($0.03/0.0027$, see the table in
figure\,\ref{fig:ptree}) times higher, or $3\cdot 10^{-2}$. These
systems, however, are probably harder to identify as post-supernova
planetary systems except maybe for the curious orbit of the single
planet.  More than 20\% of these planets have retrogade orbits
compared the their pre-suprenova orbit. This could potentially be
observed in mass transfer in the pre-supernova epoch has synchronized
the secondary's rotation.

We conclude that $\aplt 10$\,x-ray binaries in the Galaxy may still
harbor a circum-binary planet, and at most $\sim 150$ massive stars
may be orbited by a planet in a wide ($\apgt 10$\,au) and highly
eccentric ($\apgt 0.9$) orbit.  Note, however, that we assumed that
every binary is orbited by a planet, which seems unlikely (see
\cite{Kennedy2008}).

The survivability of the circum-binary planets is mediated by the
large proportion of stripped (post-mass transfer) supernovae in our
simulations. Such a type Ib/c supernova results in a relatively small
mass loss, which helps to keep the planet bound. Together with the
probability of a low kick velocity, mediates in the survivability of
planets around post-supernova systems.  The adopted distribution
functions for the kick velocity are important for this study because
of the planet's survival.  There are not many constraints to the
low-velocity tail of the supernova kick distribution
\cite{1997A&A...328L..33P}. In a future study, it may be worth
exploring this part of parameter space more exhaustively.  Finally,
our conclusions should be checked against observations.

\section*{Energy consumption of this calculation}
 
The population synthesis calculations for $4.8\times10^4$ zero-age
binaries took about 3 seconds per binary on a single Xeon E-2176M core
($\sim 12$\,Watt) resulting in about 41 hours of CPU time or 0.5\,kWh.
Building the database and repeating the calculations because of
earlier errors or tuning the selected initial conditions we should
multiply the runtime by a factor of 2 or 3.  With $0.649$\,kWh/kg
(Dutch norm for gray electricity) results in $\sim 2$\,kg CO2, which
is comparable to a daily commute.

%

\section*{Software used for this study}

In this work we used the following packages: \texttt{python}
\cite{python2,python3}, \texttt{SeBa} \cite{PortegiesZwart1996,
  Toonen2012}, \texttt{AMUSE} \cite{2018araa.book.....P},
\texttt{numpy} \cite{numpy}, \texttt{scipy} \cite{scipy},
\texttt{sklearn} \cite{sklearn}, \texttt{matplotlib}
\cite{Hunter2007}, and \texttt{sqlite3}.

\section*{Acknowledgments}

We thank the anonymous referee for spotting an error in our earlier
version of the manuscript, and for useful comments that helped us
improving the manuscript.  This work was performed using resources
provided by the Academic Leiden Interdisciplinary Cluster Environment
(ALICE).

\end{document}